\documentclass[10pt]{iopart}
\usepackage{graphicx}
\usepackage{subfig}
\usepackage{epstopdf}
\usepackage{citesort}
\newcommand{\grscale}{0.45}
\begin{document}

\title[Meas. of the magnetic field profile in the atomic fountain clock FoCS-2]{Measurement of the magnetic field profile in the atomic fountain clock FoCS-2 using Zeeman spectroscopy}

\author{
Laurent Devenoges$^1$,
Gianni Di Domenico$^2$,
Andr\'e Stefanov$^3$,
Antoine Jallageas$^2$,
Jacques Morel$^1$,
Thomas S\"udmeyer$^2$ and
Pierre Thomann$^2$
}
\address{$^1$ Federal Institute of Metrology METAS, 3003 Bern-Wabern, Switzerland}
\address{$^2$ Laboratoire Temps-Fr\'equence, Universit\'e de Neuch\^atel, 2000 Neuch\^atel} 
\address{$^3$ Institute of Applied Physics, University of Bern, 3012 Bern, Switzerland}

\ead{laurent.devenoges@metas.ch}

\begin{indented}
\item[]November 2016
\end{indented}

\begin{abstract}
We report the evaluation of the second order Zeeman shift in the continuous atomic fountain clock FoCS-2. Because of the continuous operation and of its geometrical constraints, the methods used in pulsed fountains are not applicable. We use here time-resolved Zeeman spectroscopy to probe the magnetic field profile in the clock. The pulses of ac magnetic excitation allow us to spatially resolve the Zeeman frequency and to evaluate the Zeeman shift with a relative uncertainty smaller than $1\times 10^{-16}$.
\end{abstract}

\vspace{2pc}

\section{Introduction}
\label{section0}

\begin{figure}[tbp]
\centering
\includegraphics[width=\grscale\textwidth]{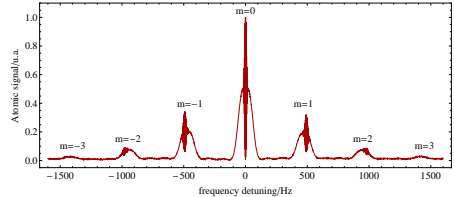}
\caption{Microwave spectrum of the seven Zeeman sub-levels. The $m\neq0$ Ramsey fringes are slighly shifted and distorded on the top of the Rabi pedestal due the magnetic field inhomogeneity.}
\label{figRamsey} 
\end{figure}

With the advent of laser cooling, thirty years ago, thermal atomic beams were progressively replaced by cold atomic fountains to contribute to the International Atomic Time (TAI) as primary frequency standards. This approach has made possible important advances in time and frequency metrology in such a way that the SI second is nowadays realized with an uncertainty below $10^{-15}$~\cite{Gerginov2009,Szymaniec2010,Guena2012}. However, state-of-the-art fountain clocks are all based on a pulsed mode of operation: the atoms are sequentially laser-cooled, launched vertically upwards and interrogated during their ballistic flight before the cycle starts over again~\cite{WynandsPaper2005}. Consequently, although the evaluation of the uncertainty budget is different for every laboratory, the methods used to measure the frequency shifts are globally the same. Our alternative approach to atomic fountain clocks is based on a continuous beam of cold atoms. Besides making the intermodulation effect negligible~\cite{JoyetPaper2001,GuenaPaper2007,DevenogesPaper2011}, a continuous beam is also interesting from the metrological point of view. Indeed, the relative importance of the contributions to the error budget is  different~\cite{DevenogesPhd2012,JallageasPaper2016}, in particular for density related effects (collisional shift and cavity pulling) and furthermore the evaluation methods differ notably. In this context, the evaluation of the second order Zeeman shift which requires a precise knowledge of the magnetic field between the two microwave interactions, in the free evolution zone, is a case in point. The methods developed in pulsed fountains to measure the magnetic field are based on throwing clouds of atoms at different heights in the resonator. This technique is not applicable to the continuous fountain since the geometrical constraints limit the range of possible launching velocities which corresponds to $\pm0.05$~m on the apogee of the nominal atomic parabola. Moreover, the atomic beam longitudinal temperature is higher in the continuous fountains ($75$~$\mu$K) than in a pulsed one ($1$~$\mu$K) and the distribution of apogees is wider. As a consequence this large distribution of transit time modifies significantly the Ramsey pattern, reducing the number of fringes as shown in Fig.~\ref{figRamsey}.\\ 
In the past, the use of Zeeman transitions $\Delta F=0$, $\Delta m=\pm1$ to probe the magnetic field was already proposed~\cite{VanierAudoinBook1989} and successfully demonstrated for thermal beams~\cite{ShirleyPaper2003}. Here we propose to adapt and to improve this technique for the evaluation of the continuous fountain clock FoCS-2~\cite{DevenogesPhd2012}. We report the use of time-resolved Zeeman spectroscopy to investigate the magnetic field in the atomic resonator where the free evolution takes place.
In Sec.~\ref{secSetup} we will give a brief description of the continuous atomic fountain FoCS-2. Then we will explain the principle of the time-resolved Zeeman spectroscopy in Sec.~\ref{secZeemanSpectro} and present the experimental results in Sec.~\ref{secExpResult}. An analysis of the numerical treatment is presented in Sec.~\ref{secSimul} and the different sources of uncertainty are discussed in Sec.~\ref{secDiscussion}.

\section{Continuous atomic fountain clock FoCS-2}
\label{secSetup}

\begin{figure}[tbp]
\centering
\includegraphics[width=\grscale\textwidth]{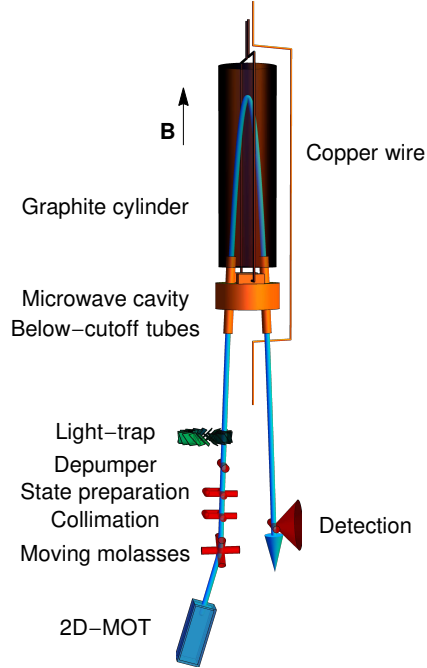}
\caption{Schematic of the continuous atomic fountain clock FoCS-2. An intense and slow atomic beam of pre-cooled cesium atoms is produced in the two-dimensional magneto-optical trap (2D-MOT). The atoms are then captured by the 3D moving molasses which further cools and launches the atoms at a speed of $4$~m/s. Then the atomic beam is collimated with Sisyphus cooling in the transverse directions, and before entering the microwave cavity, the atoms are pumped in $\left|F=3,m=0\right\rangle$ by state preparation. A light-trap located just above the last laser beam (depumper), used to pump the residual atoms in $F=4$ to $F=3$ state, prevents any light from perturbing the atoms in the interrogation zone. The atoms evolve freely for $T\approx 0.5$~s in a graphite cylinder to avoid any high frequency perturbations. Finally, after the second passage in the microwave cavity, the transition probability is measured by fluorescence detection of the atoms in $F=4$. A magnetic field is applied vertically in the free evolution zone to lift the energy degeneracy of Zeeman sublevels.}
\label{figFountainScheme}
\end{figure}

A simplified scheme of the continuous atomic fountain clock FoCS-2 is shown in Fig.~\ref{figFountainScheme}. A slow atomic beam, produced with a two-dimensional magneto-optical trap~\cite{CastagnaPaper2006}, feeds a 3D moving molasses which further cools and launches the atoms at a speed of $4$~m/s~\cite{BerthoudPaper1999}. The longitudinal temperature at the output of the moving molasses is $75$~$\mu$K. Before entering the microwave cavity, the atomic beam is collimated by transverse Sisyphus cooling and the atoms are pumped into $\left|F=3,m_\mathrm{F}=0\right\rangle$ with a state preparation scheme combining optical pumping with laser cooling~\cite{DiDomenicoPaper2010}. After these two steps, the transverse temperature is decreased to approximately $4$~$\mu$K while the longitudinal temperature stays the same. To prevent any light from entering in the free evolution zone and perturbing the atoms, a light-trap is installed just after the last laser beam~\cite{FuzesiPaper2007}. During the first passage through the microwave cavity, a $\pi/2$-pulse (duration of $8$~ms) creates a state superposition which evolves freely for approximately $0.5$~s. The Ramsey interrogation is completed by a similar $\pi/2$-interaction in the second interaction zone. Finally, the transition probability between $\left|F=3,m_\mathrm{F}\right\rangle$ and $\left|F=4,m_\mathrm{F}\right\rangle$ is measured by fluorescence detection of the atoms in $F=4$. The clock frequency is obtained by locking a local oscillator on the $\left|F=3,m_\mathrm{F}=0\right\rangle\rightarrow\left|F=4,m_\mathrm{F}=0\right\rangle$ resonance. To lift the energy degeneracy and ensure a good magnetic homogeneity, a constant magnetic field is applied throughout the free evolution zone. This field is produced by a $440$ turn solenoid ($\phi=39.5$~cm and $H=56.1$~cm) surrounded by three cylindrical magnetic shields ($\mu$-metal) plus one extra layer which embraces the whole fountain. Moreover, six extra coils ($\phi=11$~cm) located at both ends of the interrogation region limit the spatial variation of the magnetic field near the end caps. Last but not least, a copper wire running vertically aside the atomic trajectory can be used to demagnetize the four magnetic shields with a strong and low frequency current ($20$~Hz). This wire is also used to produce the excitation field in the Zeeman spectroscopy measurements described below.

\section{Zeeman spectroscopy principle}
\label{secZeemanSpectro}

The magnetic field in the free flight region lifts the degeneracy between the seven Zeeman sublevels of the $F=3$ and the nine Zeeman sublevels of the $F=4$ ground states of cesium atoms. At first order, the frequency difference between the aforementioned sub-levels ($m\times f_{z}$ with $m=0$,$\,\pm1$,$\,\pm2$,$\,\pm3$) is directly proportional to the amplitude of the magnetic field $B$, according to $f_{z}=K_{z}\times B$, where $K_{z}=3.498$~Hz/nT is the sensitivity coefficient for the $F=4$ ground state~\cite{VanierAudoinBook1989}. Therefore, the evaluation of the spatial profile of the Zeeman frequency $f_{z}$ is directly proportional to the magnetic field probed by the atoms along their ballistic flight. Basically, in order to access spatially the Zeeman frequency, we use a pulsed excitation and measure the resulting transition probability as a function of the time delay between the pulse and the detection (i.e. the spatial position of the atoms). This measurement is done in five steps. First, we prepare the atoms in $\left|F=3,m=0\right\rangle$ using a two-laser optical pumping technique~\cite{DiDomenicoPaper2010} and consecutively transfer them into the $\left|F=4,m=0\right\rangle$ clock state with a $\pi$-interaction after a first passage through the microwave cavity. Then, we drive the $\Delta m=\pm1$ transitions with a short pulse of ac magnetic excitation ($12$~ms) on the whole atomic beam present in the atomic resonator. This is done by applying a low frequency current of the order of the Zeeman frequency ($100$ to $400$~Hz) on the copper wire running vertically in the free ballistic flight region (see Fig.~\ref{figFountainScheme}). After that, the atoms remaining in $\left|F=4,m=0\right\rangle$ are transferred back into the $\left|F=3,m=0\right\rangle$ with a second microwave $\pi$-interaction after the second passage through the microwave cavity. Finally, we measure by fluorescence detection the $F=4$ total atomic population, which is therefore proportional to the $\Delta F=0,\Delta m=\pm 1$ transition probability. A precise knowledge of the atomic beam trajectory and of its timing allows us to calculate the position of atoms contributing to the signal at the moment of the pulse. This gives us a method to measure the intensity of the dc magnetic field at each position in the atomic resonator and to reconstruct its profile along the atomic beam trajectory with a worst case spatial resolution of $0.03$~m. Experimentally, this latest is related to the maximum distance traveled by the atoms during the short pulse of the ac magnetic excitation.

\section{Experimental results}
\label{secExpResult}

\subsection{Measurement of the resonance signal}

\begin{figure}[tbp]
\centering
\includegraphics[width=\grscale\textwidth]{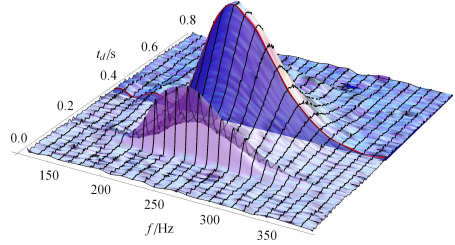}
\caption{$\Delta m=\pm1$ resonance signal as a function of the time delay $t_{d}$ measured for different values of the excitation frequency $f$ from $125$~Hz to $385$~Hz by $10$~Hz steps. The solid black lines are the measured signals, while the blue surface is a three-dimensional interpolation of the experimental data.}
\label{figPlot3D} 
\end{figure}

To evaluate the $\Delta m=\pm 1$ transition probability, we record the fluorescence signal of the $F=4\rightarrow F^{'}=5$ optical transition as a function of the time delay $t_{d}$ after the ac Zeeman magnetic pulse for different values of the excitation frequency $f$. The amplitude of the pulse ($0.7$~mA$_{\mathrm{rms}}$) and its duration ($\sim 12$~ms) are chosen to avoid any saturation of the transition (probability $<<1$) and to guarantee a sufficient spectral resolution. We show in Fig.~\ref{figPlot3D} the measured resonance signal as a function of $f$ between $125$~Hz and $385$~Hz with $10$~Hz steps and the time delay $t_{d}$. The solid black lines represent the recorded signals, while the light-blue surface is a three-dimensional interpolation of the experimental data. We highlighted in red the shape of the atomic Zeeman resonance at $t_{d}=0.4$~s as a function of $f$ (see Sec.~\ref{subsecModelExcit}).

\subsection{Model of the excitation probability and determination of the Zeeman frequency}
\label{subsecModelExcit}

\begin{figure}[tbp]
\centering
\subfloat[{}]{\includegraphics[width=\grscale\textwidth]{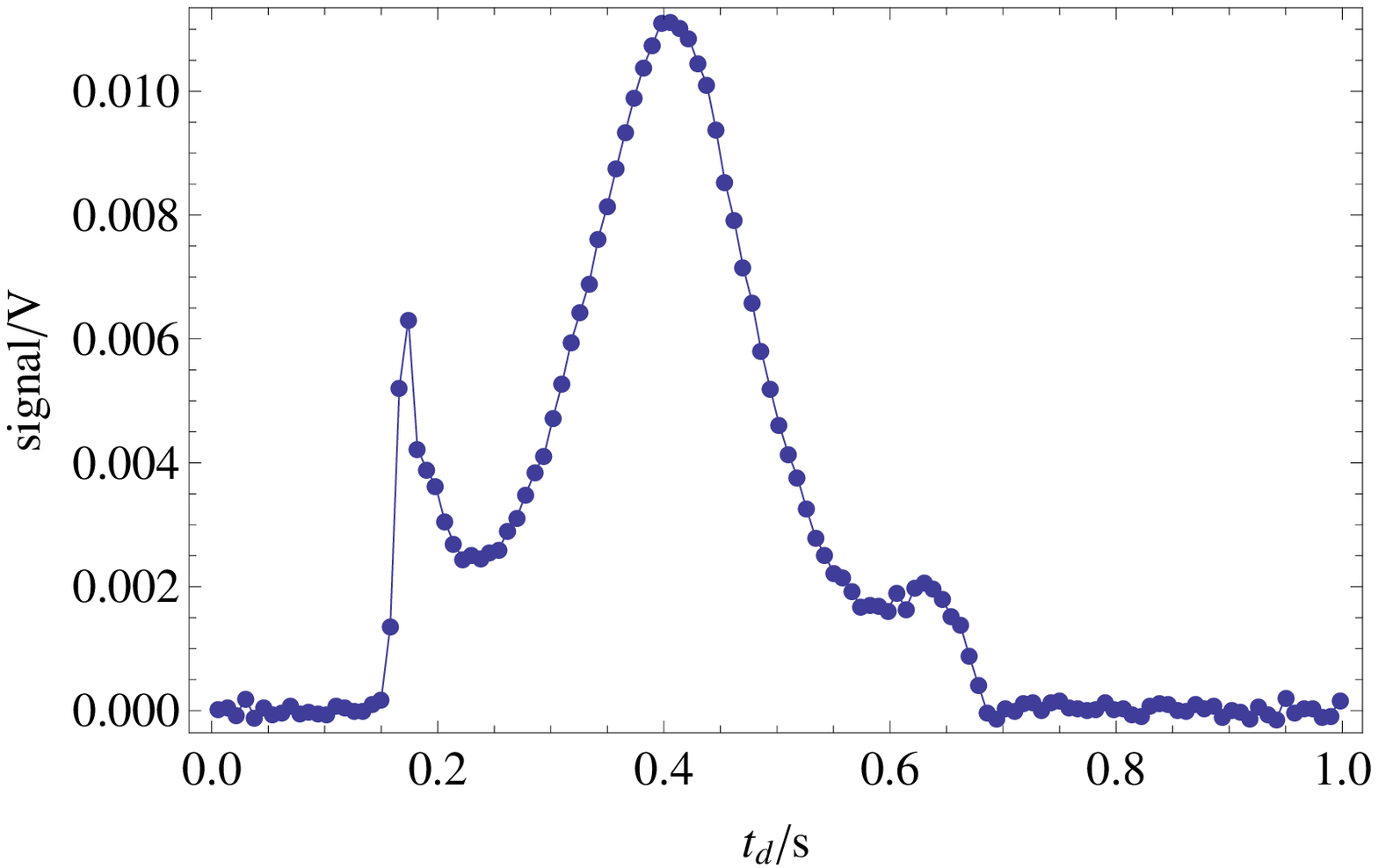}\label{figFitplota}}
\qquad
\subfloat[{}]{\includegraphics[width=\grscale\textwidth]{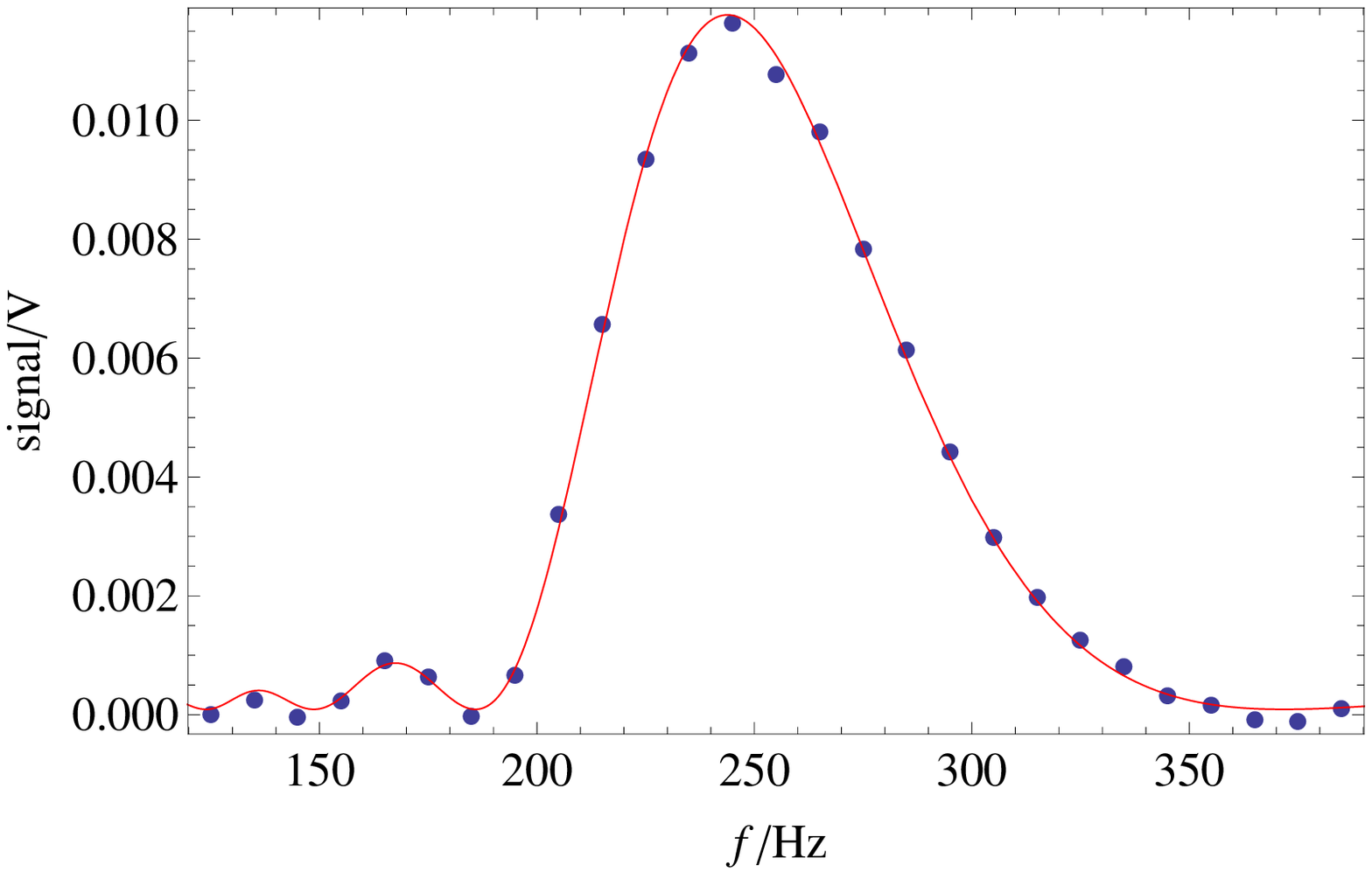}\label{figFitplotb}}
\caption{(a) Resonance signal as a function of the time delay $t_{d}$ measured for an excitation frequency $f=250$~Hz. This curve represents one of the black solid line described in Fig.~\ref{figPlot3D}. The first maxima at $t_{d}=0.2$~s is due to the high local coupling between the ac field and the atoms at the output of the microwave cavity (see Fig.~\ref{figACField} for details). (b) Example of the atomic Zeeman resonance at $t_{d}=0.4$~s after the ac low frequency pulse. The dots represent the $\Delta m=\pm1$ resonance signal as a function of the excitation frequency $f$. These data represent a transverse cut through the 3D interpolation of the Fig.~\ref{figPlot3D}. The solid red line is a fit of our excitation probability model on the experimental data. The fit parameters are the Zeeman frequency $f_{z}=247.84$~Hz, the amplitude scaling factor $a=0.0718$, and the constant offset $c=9\times 10^{-5}$.}
\end{figure}
\begin{figure}[tbp]
\centering
\includegraphics[width=\grscale\textwidth]{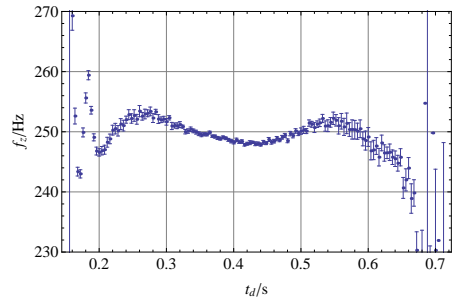}
\caption{Measured Zeeman frequency as a function of the time delay after the ac excitation pulse. Every point and its associated uncertainty  are obtained through the fit of our excitation model with the experimental data (see text for details). The Zeeman frequencies are not well defined for $t_{d}<0.155$~s. This limit corresponds to the time of flight between the microwave cavity and the detection.}
\label{figfztdplot} 
\end{figure}

The excitation pulse is given by the function:
\begin{equation} 
h(t)=\mathrm{rect}(t/\Delta t)\,\sin(2\pi ft)
\label{eqexcitationFunc} 
\end{equation}
where the $\mathrm{rect}(x)$ function is equal to 1 for $0<x<1$ and $0$ otherwise, $\Delta t$ is the pulse duration and $f$ is the excitation frequency. For a low ac amplitude, the $\Delta m=\pm 1$ transition probability is proportional to the power spectral density of the excitation pulse at a Zeeman frequency $f_{z}$: 
\begin{equation} 
P(f)=\left|\hat{h}(f_{z})\right|^{2}
\end{equation} 
where $\hat{h}$ is the Fourier transform of the excitation pulse $h$. With the excitation function above (Eq.~\ref{eqexcitationFunc}), we obtain:
\begin{equation}
P(f)=\left|\frac{2e^{-i\varphi f_{z}/f}\left(f\cos\varphi+if_{z}\sin\varphi\right)-2f}{4\pi\left(f_{z}-f\right)\left(f_{z}+f\right)}\right|^{2}
\label{eq:P1}\end{equation}
with $\varphi=2\pi f\Delta t$. Experimentally, the pulse duration $\Delta t$ is related to the number of cycles of the ac excitation $n$ and its frequency $f$. When $n$ is an integer, as it can be selected on the function generator used for this measurement, $\Delta t=n/f$ and Eq.~\ref{eq:P1} reduces to: 
\begin{equation}
P(f)=\left|\frac{n\left(\mathrm{sinc}\left(\frac{\pi n\left(f_{z}+f\right)}{f}\right)-\mathrm{sinc}\left(\frac{\pi n\left(f_{z}-f\right)}{f}\right)\right)}{2f}\right|^{2}
\label{eq:P2}
\end{equation}
In order to deduce the local Zeeman frequency $f_{z}$ from the resonance profiles, we fit the experimental curves (Fig.~\ref{figPlot3D}) with the model function $a\,P(f)+c$, where $P(f)$ is given by Eq.~\ref{eq:P2} and the only three adjustable parameters are $f_{z}$, $a$ (an amplitude factor) and $c$ (a constant offset). Fig.~\ref{figFitplota} presents the raw resonance signal as a function of the time delay $t_{d}$ and Fig.~\ref{figFitplotb} shows the experimental data presented in Fig.~\ref{figPlot3D} for a fixed time delay $t_{d}=0.4$~s (blue dots) together with the fit model (red curve). The very good agreement between the data and the analytical formula allows us to repeat this procedure for all time delays in Fig.~\ref{figPlot3D} ($0.0$~s to $0.8$~s) to determine the atomic Zeeman frequency as a function of $t_{d}$ with the associated fit parameters uncertainty $u_{f_{\mathrm{mes}}}=0.7$~Hz. The results of this analysis are shown in Fig.~\ref{figfztdplot} and give consistent values for $0.155$~s$<t_{d}<0.68$~s. These boundaries correspond to the atomic time of flight inside the C-field region regarding to the detection time.

\subsection{Determination of the spatial magnetic field profile}
\label{MagneticSpatial}

\begin{figure}[tbp]
\centering
\includegraphics[width=\grscale\textwidth]{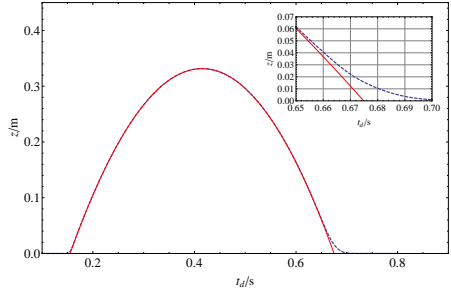}
\caption{Relation between the time delay $t_{d}$ and the average vertical position of the atoms $z$. The blue dashed line is calculated by numerical averaging of the nominal trajectory over the measured transit time distribution. The solid red line corresponds to the nominal trajectory of a mono kinetic beam.}
\label{figTzplot} 
\end{figure}

In order to determine the spatial magnetic field profile, we have to calculate the relation between the time delay $t_{d}$ and the average vertical position of the atoms $z$ at the moment of the excitation pulse. To this end, we used Fourier analysis of the Ramsey fringes~\cite{DiDomenicoPaper2011} to measure the distribution of transit times of the atoms in the interaction zone. With this method, we obtained $T=0.533$~s and $\sigma(T)=0.019$~s where $T$ is the transit time for this measurement. Note that, in order to explore the magnetic field around the nominal apogee, we used a launching velocity of $4.05$~m/s which is higher than the nominal velocity of $3.98$~m/s. The average vertical position can then be computed by averaging the nominal atomic trajectory (mono kinetic beam) over the transit time distribution. Moreover, one has to be aware that the ac magnetic excitation zone is slightly different from the Ramsey free evolution zone due to the shielding effect of the microwave cavity. If we define $z=0$~mm to be the altitude above which ac excitation of the atoms is detectable, then the microwave cavity center corresponds to $z_{c}=-15$~mm and the Ramsey free evolution zone, according to~\cite{JoyetPhd2003}, begins at $z_{0}=-18$~mm. The result of the averaging is shown in Fig.~\ref{figTzplot}. 

\begin{figure}[tbp]
\centering
\centering
\subfloat[{}]{\includegraphics[width=\grscale\textwidth]{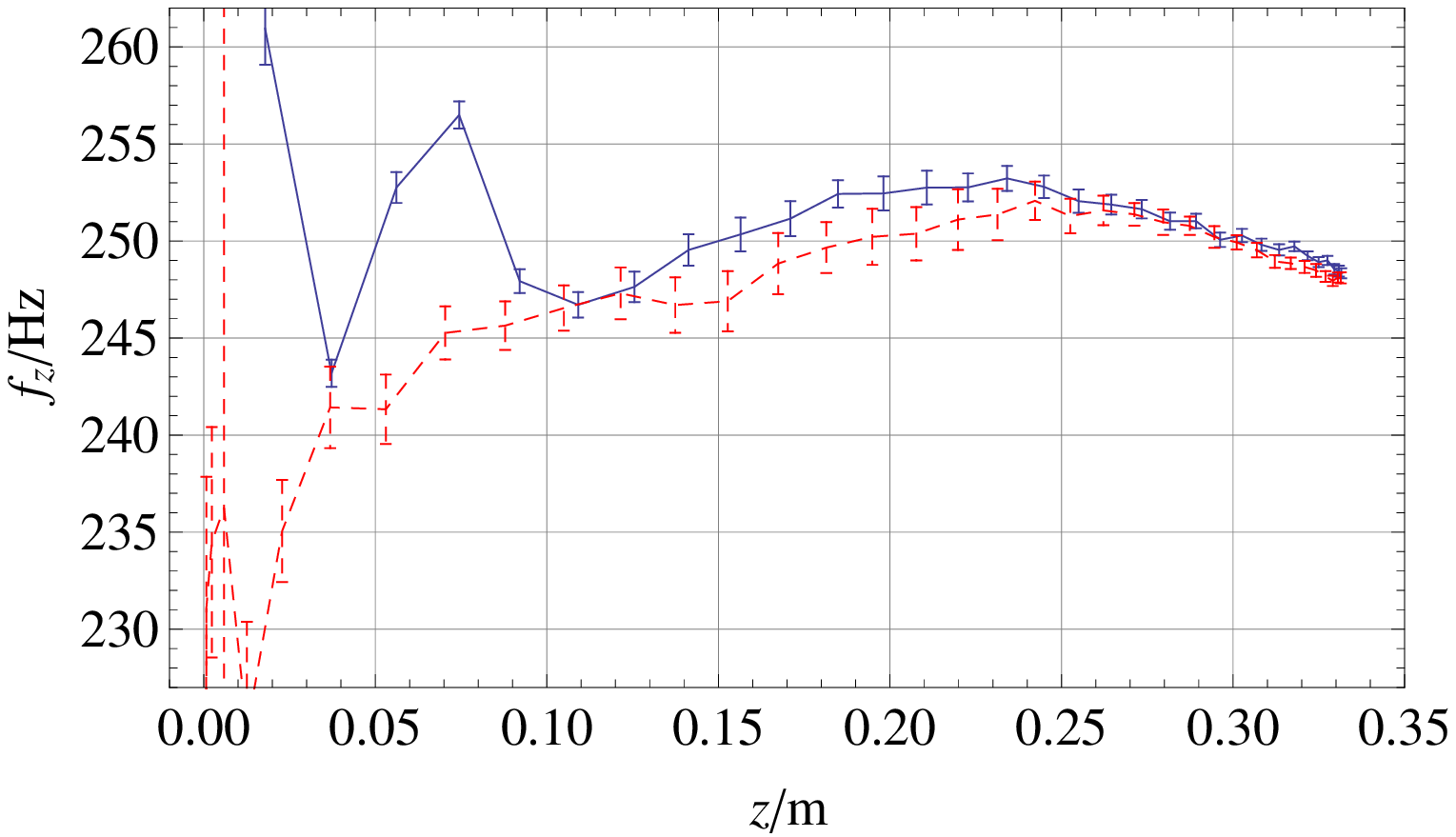}\label{figfzupdnplot}}
\qquad
\subfloat[{}]{\includegraphics[width=\grscale\textwidth]{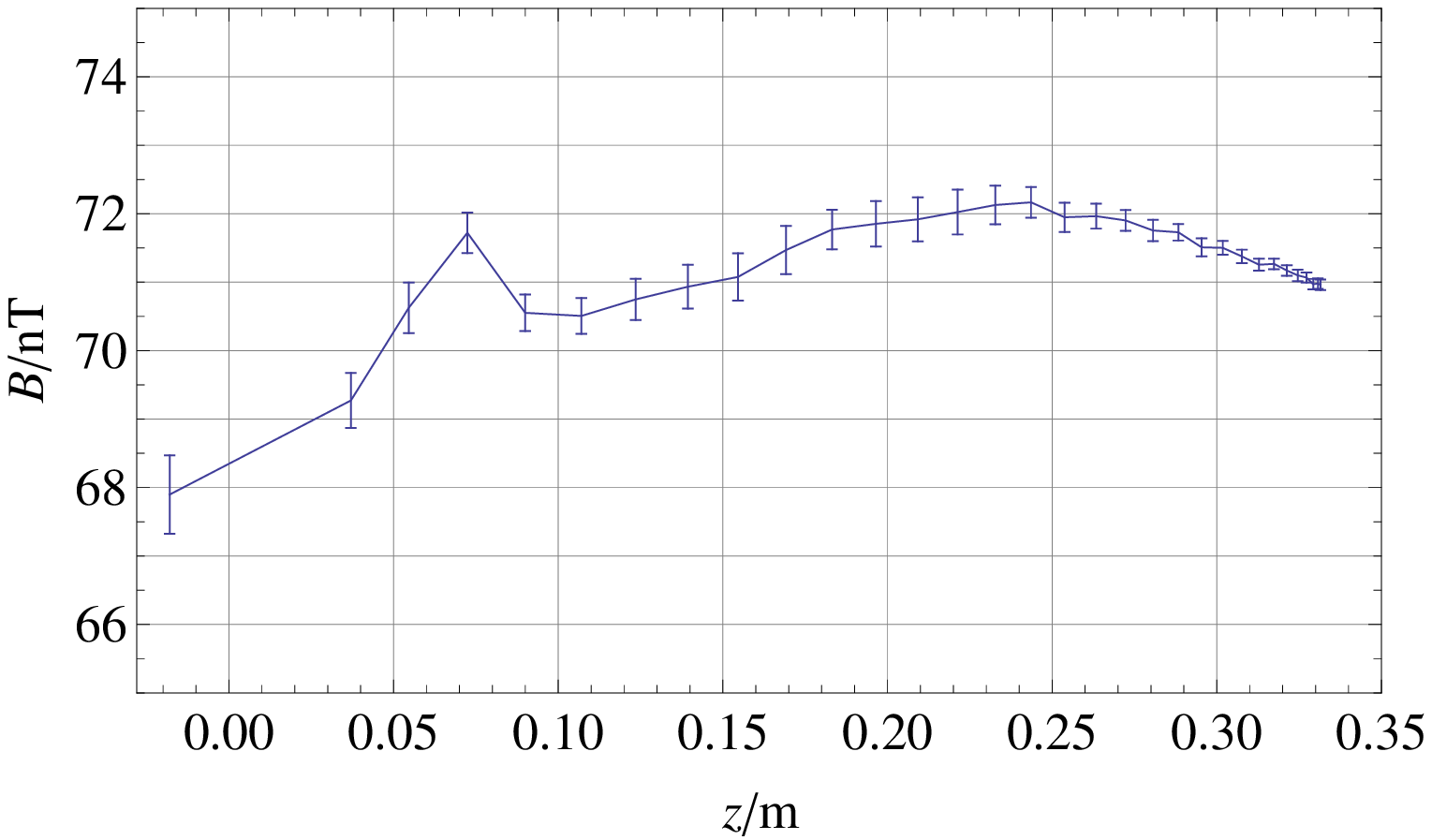}\label{figBzmplot}}
\caption{(a) Zeeman frequency as a function of the vertical position $z$ calculated from the time-resolved measurement of the Zeeman frequency shown in Fig.~\ref{figfztdplot}. The blue (red dashed) lines correspond to atoms going down (up) at the moment of the excitation pulse. Note: $z=0$~mm is defined to be the altitude above which the atoms feel the ac excitation pulse i.e. $15$~mm above the microwave cavity center. (b) Averaged spatial magnetic field profile probed by the atoms $(f_{\mathrm{up}}+f_{\mathrm{dn}})/2$ (see text for details). To measure the magnetic field at $z_{0}=-18$~mm we used the position of the $m=-1$ Rabi pedestal of the microwave spectrum.}
\end{figure}

This numerical function allows us to compute the Zeeman frequencies along the vertical trajectory. This spatial profile is shown in Fig.~\ref{figfzupdnplot}. 

As we can see, the Zeeman frequencies are almost identical for atoms going up and down above $z=0.1$~m. The difference below $z=0.1$~m is due to a component of the fixed cavity cradle which creates a magnetic field inhomogeneity (this assumption has been verified by turning the cavity by $180^{\circ}$). Because of the evaluation of the second order Zeeman shift only requires the knowledge of the average of the magnetic field probed by the atoms on the parabolic trajectory (see Sec.~\ref{secDiscussion}), we calculated the average between the Zeeman frequency probed on the way up ($f_{\mathrm{up}}$) and on the way down ($f_{\mathrm{dn}}$) with $(f_{\mathrm{up}}+f_{\mathrm{dn}})/2$ and divided the result by $K_{z}(F=4)=3.498$~Hz$/$nT to obtain the magnetic field profile shown in Fig.~\ref{figBzmplot}. Note that this averaging leads to an additional uncertainty which has been numerically evaluated in Sec.~\ref{secSimul}. The value of the magnetic field at $z=-18$~mm was obtained with the position of the $m=-1$ Rabi pedestal of the microwave spectrum.


\subsection{Determination of the time-averaged magnetic field}
\label{sec4}

\begin{figure}[tbp]
\centering
\includegraphics[width=\grscale\textwidth]{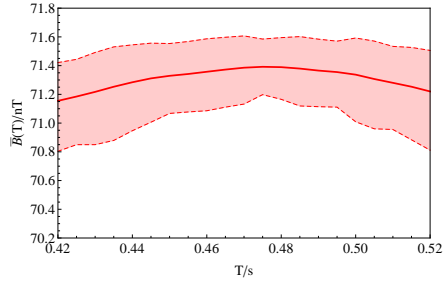}
\caption{Time-averaged magnetic field along the atomic trajectory as a function of the transit time $T$. The solid red line has been determined by numerical integration of the spatial profile of the magnetic field $B(z)$ shown in Fig.~\ref{figBzmplot}. The red dashed lines represent the $\pm1\sigma$ error band. They were calculated using the propagation of the errors in eq.~\ref{integrB} (see text for details).}
\label{figBmplot} 
\end{figure}

We determined the time-averaged of the magnetic field $\overline{B}(T)$ seen by the atoms during their free evolution by numerical integration of the spatial magnetic field profile $B(z)$ shown in Fig.~\ref{figBzmplot}. We calculated $\overline{B}(T)$ as follows: 
\begin{equation}
\overline{B}(T)=\frac{1}{T}\int_{0}^{T}B\left(z(t,T)\right)dt
\label{integrB}
\end{equation}
where $z(t,T)=g\, t\,(T-t)/2+z_{0}$ is the trajectory of the atomic beam and $g=9.81$~m$/$s$^{2}$. We repeated this procedure for different values of the transit time around the nominal value of $T=0.512$~s to represent graphically the temporal variation of the magnetic field $\overline{B}(T)$. The result is shown in Fig.~\ref{figBmplot} together with the $\pm1\sigma$ error band. This band is calculated using the same analysis and the propagation of the errors in eq.~\ref{integrB} by considering the uncertainty on the magnetic field measurement $u_{f_{mes}}$ (see Sec.~\ref{subsecModelExcit}) and the uncertainty on the transit times distribution $u_{\rho(T)}$ (see Sec.~\ref{MagneticSpatial}). At $T=0.512$~s we obtain $\overline{B}(0.512)=71.2$~T and $u_{B_{\mathrm{mes}}}=0.2$~nT.

\section{Simulation}
\label{secSimul}

In order to validate the method and the experimental analysis, we perform numerical simulation and estimate the uncertainty due to the approximation in the relation between the position of the atoms and the extraction of the resonance frequency. For a given Zeeman frequency dependency on $z$, we numerically compute the expected transition probability as a function of the time delay and of the excitation frequency. From those simulated data, we extract a Zeeman frequency dependency as we have done in the experiment. Finally we compare the estimated Zeeman frequency with the original one, the difference giving us an estimate of the experimental uncertainty. 

\begin{figure}[tbp]
\centering
\includegraphics[width=0.35\textwidth]{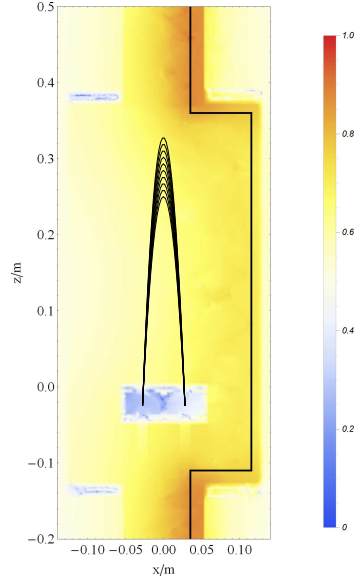}
\caption{Relative amplitude $A(x,z)$ of the low frequency ac electromagnetic field in the plane of the atomic trajectories produced by the vertical copper wire running in the free evolution zone. We can see the variations of the field probed by the atoms on the way down which is due to the close proximity of the copper wire.}
\label{figACField} 
\end{figure}

\begin{figure}[tbp]
\centering
\includegraphics[width=\grscale\textwidth]{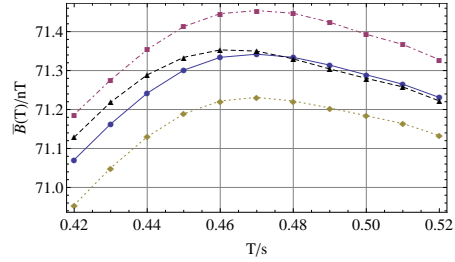}
\caption{Simulated time average magnetic field along the atomic trajectory as a function of the transit time $T$ for the ac field shown in Fig.~\ref{figACField}. The purple squares (yellow diamond) correspond to atoms going down (up) and the blue dots are the average of both. The black triangles are the given time average field dependency used for the simulation.}
\label{figBTsimulwithamplitude} 
\end{figure}

In a first step, we numerically simulate the amplitude of the ac magnetic field in the plane of the atoms trajectories. As mentioned previously, the field is created by the copper wire located in the vacuum chamber. Using an finite element algorithm for the known geometry, we compute the amplitude of the ac field $A(x,z)$ in the whole atomic resonator as shown on Fig.~\ref{figACField}. To determine the transition probability as a function of the detection time $t_{d}$ and the applied ac excitation frequency $f$, we would have to compute the transition probability from the time dependent Schr\"odinger equation. For simplicity reasons, we use here a more basic model. We define the transition probability for an atom with a time of flight $T$ and subjected to a weak ac field $A(x,z)$ with a square pulse shape by:
\begin{equation}
p\left(f,T\right)=\left|\mathrm{sinc}\left(2\pi\frac{f-f_{z}\left(z\right)}{2}\Delta t\right)\right|^{2}\times\left|A\left(x,z\right)\right|^{2}
\end{equation}
where $\Delta t$ is the pulse duration. Finally the total signal at the detection is computed by integrating this probability over the time where the ac pulse is on and averaging over the time of flight distribution $F\left(T\right)$:
\begin{eqnarray}
 \eqalign{p\left(t_{d},f\right) &= \int\int \mathrm{dt}\,\mathrm{dT}\,F\left(T\right)\mathrm{rect}\left(t/\Delta t\right)\\
  &\times\left|\mathrm{sin}\left(2\pi\frac{f-f_{z}\left(z\left(t,T\right)\right)}{2}\Delta t\right)\right|^{2}\\
  &\times\left|A\left(x\left(t,T\right),z\left(t,T\right)\right)\right|^{2}}
\end{eqnarray}
From the resulting function $p\left(t_{d},f\right)$, an estimation of the Zeeman field is obtained with the same treatment as the one used for the experimental data (see Sec.~\ref{subsecModelExcit}). The resulting dependency of $\overline{B}(T)$ is shown in Fig.~\ref{figBTsimulwithamplitude} together with the original given ac Zeeman field. The estimated uncertainty on $\overline{B}(T)$ due to this analysis is measured as the difference between the original field (black triangles) and the one computed with our method (blue dots). The numerical treatment adds an error on the time-averaged magnetic field of $10$~pT.

Furthermore, to estimate the contribution to the uncertainty due to the non-homogeneous ac field, we repeated the same simulation with a uniform ac field in the free evolution zone (not shown here). The difference between the two results for $T=0.512$~s differs by $0.1$~nT. By summing quadratically these two contributions, we compute that the numerical analysis performed on the experimental data adds a total uncertainty on the time-averaged magnetic field $u_{B_{\mathrm{num}}}=0.1$~nT.

\section{Discussion}
\label{secDiscussion}

The second order Zeeman shift in atomic clocks scales with the time average of the square of the magnetic field probed by the atoms in the free evolution zone. This effect is calculated with~\cite{VanierAudoinBook1989}:

\begin{eqnarray}
\eqalign{\Delta\nu_{z} & =\nu_{0}(B)-\nu_{0}=K_{0}\,\overline{B^{2}}\\
& = \frac{1}{2\nu_{0}}\frac{(g_{\mathrm{J}}+g_{\mathrm{I}})^{2}\mu^{2}_{\mathrm{B}}}{h^{2}}\overline{B^{2}}}
\label{eqZS}
\end{eqnarray}

where $\nu_{0}$ is the unperturbed cesium frequency, $g_{\mathrm{J}}$ and $g_{\mathrm{I}}$ are the Land\'e factors of the angular momentum for the electron and the nucleus, $\mu_{\mathrm{B}}$ is the Bohr magneton, $h$ is the Planck constant and $\overline{B^{2}}$ is the time average of the square of the magnetic field probed by the atoms between the microwave interactions. We discuss now how the magnetic field uncertainties affect the evaluation of the second order Zeeman shift. For this purpose, we decompose the total measurement uncertainty of $B(T)$ in two contributions.\\

In Sec.~\ref{secExpResult} we saw that this time-resolved Zeeman spectroscopy technique gives a direct access to the spatial magnetic field profile in the free evolution zone. However to evaluate the time-averaged magnetic field probed by the atoms and the second order Zeeman shift, it is necessary to compute the integral (eq.~\ref{integrB}). By definition, this calculation give $\overline{B}(T)$ and thus the resulting error using $\left(\overline{B}(T)\right)^{2}$ instead of $\overline{B^{2}}(T)$ must be considered:

\begin{equation}
u_{\sigma^{2}_{B}}=K_{0}\,\sigma^{2}_{B}(T)
\end{equation}

where $\sigma^{2}_{B}(T)=\overline{B^{2}}(T)-\left(\overline{B}(T)\right)^{2}$ represents the variation of the magnetic field $B(\textbf{r}(t))$ along the atomic trajectory. In our case we use half of the difference between the minimum and the maximum magnetic field probed by the atoms. The results shown in Fig.~\ref{figBzmplot} gives us $\sigma_{B}(T)=2$~nT, which corresponds to an uncertainty $u_{\sigma^{2}_{B}}=2\times 10^{-7}$~Hz.

The second contribution is related to the uncertainties of the magnetic field measurement $u_{B_{\mathrm{mes}}}=0.2$~nT (see Sec.~\ref{sec4}) and the uncertainty coming from the numerical treatment $u_{B_{\mathrm{num}}}=0.1$~nT discussed in Sec.~\ref{secSimul}. To be complete we also add the uncertainty on the temporal stability of the magnetic field $u_{B_{\mathrm{in}}}$ which may result from any variations of the ambient field. We evaluated this term by locking the clock on a magnetic field sensitive transition $\left|F=3,m_\mathrm{F}=1\right\rangle\rightarrow\left|F=4,m_\mathrm{F}=1\right\rangle$ for 5 days. Over this period of time, we recorded small short terms fluctuations $<0.01$~Hz with a frequency drift of $0.1$~Hz. Extrapolating this drift over one month gives us an additional conservative uncertainty $u_{B_{\mathrm{in}}}=0.2$~nT. Note that in the near future, we plan to repeat the present analysis to evaluate the long term evolution of the magnetic field. The quadratic sum of these contributions allows us to calculate the uncertainty on the frequency: 

\begin{equation}
u^{B}_{\Delta\nu_{z}}=2K_{0}\,\overline{B}(T)\,\sqrt{u^{2}_{B_{\mathrm{mes}}}+u^{2}_{B_{\mathrm{num}}}+u^{2}_{B_{in}}}
\end{equation}

which leads to $u^{B}_{\Delta\nu_{z}}=1.9\times 10^{-6}$~Hz for the measured magnetic field at the optimal transit time $\overline{B}(0.512)=71.2$~nT.

Finally, we compute the total uncertainty for the second order Zeeman shift as the quadratic sum of the two terms described above: 

\begin{equation}
u_{\Delta\nu_{z}}=\sqrt{(u^{B}_{\Delta\nu_{z}})^{2}+(u_{\sigma^{2}_{B}})^{2}}=1.9\times 10^{-6}\,\mathrm{Hz}
\label{eqErZS}
\end{equation}

\section{Conclusion}
\label{secConclusion}

In this paper we presented the evaluation of the second order Zeeman shift in the continuous fountain FoCS-2 using time-resolved Zeeman spectroscopy method to probe the Zeeman frequency along the atomic trajectories. This technique allowed us to determine a time averaged magnetic field of $71.2$~nT with an uncertainty of $\pm0.3$~nT, which corresponds to a frequency shift:

\begin{equation*}
\frac{\Delta\nu_{z}}{\nu_{0}}=(23.6\pm 0.2)\times 10^{-15}
\end{equation*}

This new measurement paves the way for the complete evaluation of the continuous fountain FoCS-2.

\ack{This research was supported by the grant $200021\,141338$ of the Swiss National Science Foundation (SNF).} 

\section*{References}

\bibliography{biblioZeeman}
\end{document}